\documentclass[prd, twocolumn, nofootinbib, floatfix]{revtex4-2}
\usepackage{amsmath}
\usepackage{graphicx}
\usepackage{dcolumn}
\usepackage{bm}
\usepackage{epsfig}
\usepackage{amssymb,latexsym,mathrsfs}
\usepackage{graphicx}
\usepackage{color}
\usepackage{hyperref}

\usepackage{float}

\hypersetup{
    colorlinks=true,
    linkcolor=red,
    citecolor=blue,
}

\newcommand{\be}{\begin{equation}}
\newcommand{\ee}{\end{equation}}
\newcommand{\bs}{\begin{split}} 
\newcommand{\bea}{\begin{eqnarray}}
\newcommand{\eea}{\end{eqnarray}}

\newcommand{\om}{\Omega_m}

\newcommand{\scrm}{\mathcal{M}}

\begin{document}

\title{A Sparse Spectroscopic Supernova Survey} 

\author{Eric V.\ Linder$^{1,2}$} 
\affiliation{${}^1$Berkeley Center for Cosmological Physics \& Berkeley Lab, 
University of California, Berkeley, CA 94720, USA\\
${}^2$Energetic Cosmos Laboratory, Nazarbayev University, 
Nur-Sultan 010000, Kazakhstan
}

\begin{abstract} 
Supernova cosmology surveys are traditionally time consuming, especially 
for the critical spectroscopic data.  However, a single spectrum 
at maximum light may provide accurate distance estimation if recent 
developments hold. This could open up a new 
type of supernova cosmology survey, with a useful interaction between 
the spectra and  
a focus on specific redshifts. We optimize the redshift selection and 
show that this condensed survey could efficiently deliver highly accurate 
dark energy constraints. 
\end{abstract}

\date{\today} 

\maketitle

\section{Introduction}

Supernovae type Ia (SN) are the most incisive dark energy probes, with a 
better sensitivity to dark energy equation of state properties with respect 
to matter density than other measurements. Their distance-redshift relation 
relies predominantly on individual object measurements, substantially 
avoiding systematics issues such as survey inhomogeneity (masking and 
depth variations), source blending (SN are transients and 
have a unique spectrum), image distortion from atmosphere/telescope/detector, 
etc. SN do have their own systematics, but many of these 
can be avoided by detailed spectrophotometric measurement rather than 
a purely imaging survey. 

A significant recent development has been the improvement of SN distance 
calibration through detailed spectral information \cite{boone1,boone2}. 
Moreover, this enables an important advance in obtaining this information 
from only a single spectrum measurement at an epoch near maximum 
light. This would decrease the observational expense of obtaining high 
accuracy SN distances, while reducing both statistical dispersion and 
systematics uncertainty. 

If this promise holds, it could enable a revolution in the approach to 
supernova cosmology. A corollary impact could be in the cosmic survey 
design, to make it even more efficient. While traditionally one measures 
SN distances over a continuous redshift range, e.g.\ $z=[0,1]$, the 
kernel for cosmological parameter sensitivity is broad enough that a 
sparse survey in selected redshift slices could deliver quite accurate 
constraints. The interaction between the redshift and the observer 
frame wavelength of SN spectra could also simplify the instrumentation 
as they can imply specific wavelength 
ranges of interest. 

Section~\ref{sec:slice} introduces the method for computing the optimal 
redshift slices and evaluates the impact on the cosmological constraints. 
In Section~\ref{sec:survey} we discuss some of the ingredients necessary 
for enabling such a survey, in particular ground-space complementarity. 
We conclude in Section~\ref{sec:concl}.

\section{Cosmology in Slices} \label{sec:slice} 

The distance-redshift relation contains information on the energy 
density contents of the universe, and its evolution. As the dark energy 
density evolution depends on its equation of state $w(z)$, well measured 
distances are an excellent probe of $w(z)$, accurately characterized through 
$w(z)=w_0+w_a z/(1+z)$, where $w_0$ is the present value and $w_a$ a 
measure of its time variation. In addition, the fractional matter density 
with respect to the critical density, $\om$, enters as well as a calibration 
parameter. For SN, distances are measured with respect to the low 
redshift expansion and that parameter is $\scrm$, insensitive to 
cosmology; for a distance indicator such as baryon acoustic oscillations 
the distance is measured with respect to the high redshift universe 
and the parameter will involve $\om$, weighting the sensitivity to 
matter relative to dark energy more than for SN. That is, SN are more 
sensitive to dark energy relative to matter than other distance indicators. 

Thus, for SN the parameter set is $\{\scrm,\om,w_0,w_a\}$ and we can 
estimate the constraints on these parameters from a set of magnitude 
measurements $m(z_i)$ through the information matrix formalism. Here 
the magnitude 
\be 
m=5\log[H_0d(z)]+\scrm\,;
\ee 
note $H_0d$ is independent 
of $H_0$, depending only on $\{\om,w_0,w_a\}$. We assume a flat universe 
and set $c=1$. 

The redshift distribution $\{z_i\}$ of the set of measurements has traditionally 
been determined by the experiment characteristics, e.g.\ the limiting 
magnitude, detector properties, observing time, etc. Furthermore, since 
the occurrence of SN cannot be predicted, the surveys are generally rolling 
searches, scanning an area of the sky repeatedly with some cadence. 
However, the advent of wide, deep, time domain imaging surveys such as 
Rubin Observatory's Legacy Survey of Space and Time (LSST) combined with 
the spectral developments of \cite{boone1,boone2} derived from 
high accuracy spectrophotometric Nearby Supernova Factory \cite{nsf} data, 
can break this restriction and allow greater freedom in determining the 
target SN 
distribution from other considerations -- such as optimizing the 
cosmological leverage. We discuss some practicalities of this in 
Section~\ref{sec:survey} but first assess its possible implications. 

The leverage on dark energy properties can be compactly discussed in 
terms of the figure of merit, 
\be 
{\rm FOM}\equiv 1/\sqrt{\det[COV(w_0,w_a)]}\,, 
\ee 
where $COV$ is the inverse of the information matrix, marginalized over the other 
cosmological parameters, and we include a Planck prior on the distance to 
the cosmic microwave background distance to last scattering. 
For maximal leverage and breaking of covariances 
between parameters, the SN survey should make measurements at the lowest 
and highest redshifts available, but not necessarily all those in between. 
We set $z_{\rm low}=0.05$ and $z_{\rm max}=1$, later considering variation 
of $z_{\rm max}$. To maximize the efficiency of the survey, we want the 
minimum number of redshift slices (and hence wavelength regions), which 
is equal to the number of parameters being constrained, in this case four. 
Thus we have two free survey parameters, the redshift centers of slices 
$z_2$ and $z_3$. When observationally practical, such sparse redshift 
surveys can deliver tighter parameter estimation than covering the full range 
given resource constraints -- see \cite{huttur01,fhlt,lin15} -- and be 
highly efficient. Cosmologically, this works because the redshift kernel 
of distance sensitivity to cosmological parameters has a width of an 
appreciable fraction of the evolution, whether from an e-fold of expansion 
or other physical dynamics \cite{comb}. 

We optimize $z_2$, $z_3$ to deliver a maximum FOM, i.e.\ the most 
incisive information on dark energy properties. For the measurement 
precision we take it to be limited by some residual 
systematics for a future high accuracy spectroscopic survey. We adopt 
the Linder-Huterer (LH) prescription \cite{0208138} for the form, using 
\be 
\frac{\delta d_{\rm sys}}{d}=0.0017(1+z)\,, \label{eq:sys} 
\ee 
increasing with redshift. This 0.17\%--0.34\% accuracy 
is equivalent to 3.7 mmag at $z=0$ to 
7.4 mmag at $z=1$ -- certainly challenging, but we are aiming for a 
spectroscopic SN survey in the LSST era, using the recent machine 
learning ``Twins Embedding'' technique of \cite{boone1,boone2} 
or future further improvements. Twins Embedding is not only 
powerful on systematics mitigation but reduces the dispersion 
remaining after the spectral fit, hence a systematics limited survey 
seems reasonable (see Section~\ref{sec:survey} for a more 
quantitative discussion). We later explore the impact 
of varying this level. The LH systematic is taken to 
be coherent over a redshift slice of width $\Delta z=0.1$, so we 
impose $z_3\ge z_2+0.1$. 

\begin{figure}[!htb]
\centering 
\includegraphics[width=\columnwidth]{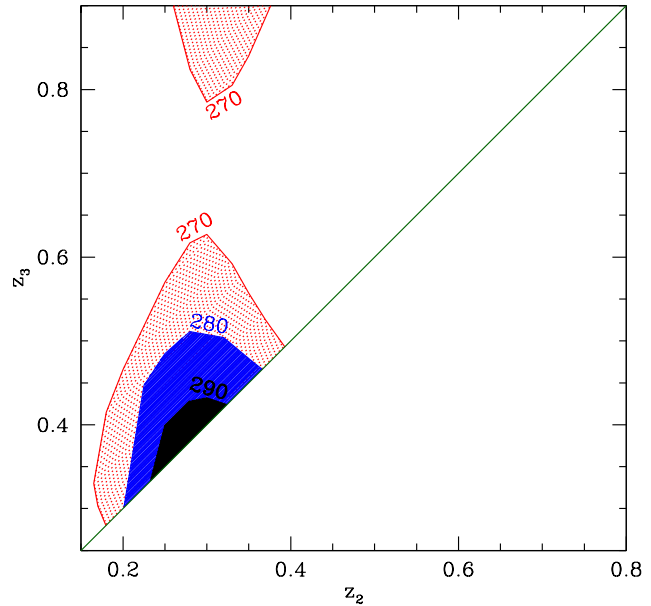} 
\caption{
As we vary the intermediate redshift slice locations $z_2$, $z_3$ of the SN survey, 
the dark energy figure of merit changes, as shown by the isocontours 
of FOM. The lower triangular region does not obey the condition 
$z_3\ge z_2+0.1$ (the equality is given by the green diagonal line) 
and is moot. The blank region in the upper triangular area has 
${\rm FOM}<270$. 
The maximum ${\rm FOM}=295$ occurs at $z_2=0.28$, $z_3=0.38$. 
} 
\label{fig:fomshade} 
\end{figure}

Figure~\ref{fig:fomshade} shows the results for how the dark energy 
figure of merit varies over the $z_2$--$z_3$ plane. The optimum is at 
$z_2=0.28$, $z_3=0.38$ with maximum FOM=295. This is quite impressive, 
comparable to many other Stage 4 dark energy experiments, even those 
with multiple probes. Of course it depends on the ability to realize 
the tight systematics control.

The excellent FOM occurs despite the sparseness of the survey, using 
only four redshift slices (of width $\Delta z\lesssim 0.1$) 
at $z=0.05$, 0.28, 0.38, 1. The exact positions of the intermediate slices 
comes from an interplay between cosmological sensitivity and systematics; 
the power of an optimized sparse survey can be seen by noting that using 
five redshift slices (so roughly 25\% more statistics) but more evenly spread 
at $z=0.05$, 0.25, 0.5, 0.75, 1 would give less than a 2\% improvement in FOM. 
Dark energy properties are constrained to $\sigma(w_0)=0.051$, 
$\sigma(w_a)=0.23$, and $\sigma(\om)=0.0055$. 
This would be a significant advance 
in our knowledge of dark energy. 

The low and high redshift limits of the SN survey are quite important. 
Raising the lowest slice to $z_{\rm low}=0.1$ would reduce the FOM by 
28\%. Lowering the highest slice to $z_{\rm max}=0.9$ reduces the FOM 
by 8\%, while raising it to $z_{\rm max}=1.1$ increases the FOM by 6\%. 
The systematics level also has significant impact: if we double its level 
then the FOM decreases by a factor 3.2, if we put a floor on the systematics 
of Eq.~\eqref{eq:sys} of 5 mmag (0.23\% in distance, 
effectively raising the systematics on 
the SN at $z_{\rm low}$ and $z_2$ to the level at $z_3$) the FOM is reduced by 13\%.

\section{Survey Considerations} \label{sec:survey} 

The previous section sets an impactful goal that future surveys can strive to 
achieve. As mentioned, one promising approach comes from the results of 
\cite{boone1,boone2} that find that a single spectrum near maximum 
light can serve as an accurate SN distance indicator. They also raise 
the intriguing point that a particular focused spectral region, between 
6600--7200 \AA\ in the supernova restframe, has especially low dispersion. 

Let us use this as a guide for exploring future possibilities. We refer 
to such a notional supernova cosmology survey 
as the Sparse Spectroscopic Supernova Survey, or $S^4$; it is sparse 
in the sense of focusing on select redshift slices and a single spectrum 
near maximum light.   
If the spectral calibration 
does fulfill its promise, then the distance standardization per SN is 
0.101 mag, or 0.073 mag with peculiar velocity contributions removed and 
an improved reference sample, while noting that the intrinsic dispersion 
found in the focused spectral region is a remarkable 0.02 mag \cite{boone1,boone2}. 
These 
three numbers correspond to 4.7\%, 3.4\%, or 0.92\% fractional distance 
uncertainty respectively. Table~\ref{tab:nsys} shows the number of SN 
(and hence spectra) needed for the statistical uncertainty to match the 
systematics level assigned in Eq.~\eqref{eq:sys}, that is 
\be 
N_{\rm sys}({\rm dispersion};z)=\left(\frac{{\rm dispersion}}{{\rm systematic}(z)}\right)^2\,. 
\ee

\begin{table}
\begin{center}
\begin{tabular}{l | c | c | c | c}
\hline
Redshift & \ 0.05\ \ & 0.28 & 0.38 & 1.0 \\
Wavelength ($\mu$m)\ \ & \ 0.76\ \ & \ 0.92\ \ & \ 0.99\ \ & \ 1.44\ \ \\ 
$N_{\rm sys}$ (0.101 mag)\ & \ 675 & 454 & 391 & 186 \\ 
$N_{\rm sys}$ (0.073 mag)\ & \ 353 & 238 & 204 & \ 98 \\ 
$N_{\rm sys}$ (0.02 mag)\ & \ \ 27 & \ 18 & \ 16 & \ \ 8 \\ 
\hline
\end{tabular}
\end{center}
\caption{The optimized $S^4$ targets supernovae at four redshift slices, 
given by the first row, with the spectrum maximum wavelength in 
the observer frame given in the second row. The third through fifth rows 
show the number of SN required for statistical uncertainty to fall below 
the systematics goal, for three choices of the residual intrinsic dispersion. 
} 
\label{tab:nsys}
\end{table}

Such a powerful survey could conceivably be accomplished with some 
1700 SN spectra. (Note that a higher systematics floor implies an even 
lower $N_{\rm sys}$.) From the maximum wavelength corresponding to 
7200 \AA\ restframe, the end of the lowest dispersion spectral range, 
we see that the lowest three redshift slices can all be surveyed using 
CCDs out to 1 $\mu$m\footnote{The 
results of \cite{boone1,boone2} use 
a restframe range of $\sim3300-8600$ \AA; it is not clear how much of 
this is crucial (dust constraints in general improve the longer the 
wavelength baseline) or whether one gains most of the impact already 
by 7200 \AA, say. In addition, germanium CCDs under 
development may allow good measurement out to 1.4 $\mu$m, 
sufficient to reach 1 $\mu$m SN restframe for the lowest three 
redshift slices \cite{germ1,germ2}.}.  
The highest 
redshift slice, at $z=1$, requires accurate near infrared (NIR) 
measurements; this will require space observations. Possibly one could 
push the highest slice to greater $z_{\rm max}$, up to $z=1.36$ for 
measurements out to 1.7 $\mu$m, and do correspondingly better on dark 
energy FOM (for a reoptimized $z=\{0.05,0.31,0.41,1.36\}$, ${\rm FOM}=344$), 
but we stay with our fiducial $z_{\rm max}=1$. 

Other observations needed for this supernova cosmology program, i.e.\ 
before targeting them with spectroscopy, are: 
finding the SN, establishing them as likely Type Ia, estimating the time 
of maximum light, and measuring the host galaxy redshift. Time domain 
imaging surveys such as LSST will find hundreds of thousands of SN and 
can use its six wavelength band observations to estimate type 
(see, e.g., \cite{avocado} and related vegetarian developments) and time of maximum. 
LSST and other galaxy catalogs can provide estimates of host galaxy redshifts. 
Remember, we only need $\sim1700$ out of the hundreds of thousands of SN 
so we are free to discard candidates lacking certainty and precision. 
The focus would be on the 
optimized redshift slices, i.e.\ SN in $z=[0.03-0.1]$, $[0.23,0.33]$, 
$[0.33,0.43]$, and $[0.95,1.05]$, say. In the coming era, accurate 
spectroscopic redshifts for a few thousand host galaxies will also be easy 
as a supplement to imaging and catalogs. 

The use of specific redshift slices can also reduce systematics by 
itself -- i.e.\ certain redshifts may be less susceptible to systematics 
by virtue of the cosmology dependence or wavelength band 
characteristics, as in the studies of k-corrections \cite{kimk}, and filter 
zero point calibration errors and population evolution \cite{samsing}. At the 
least, the systematics could be less diverse, potentially making control 
or correction easier. 

One of the difficulties with pushing 
beyond $z_{\rm max}\approx1$ is the challenge of obtaining accurate 
measurements there from ground based imaging. However, if all we need is 
typing and estimate of maximum light, this may be possible, especially as 
time dilation helps with cadenced observations at these redshifts. We 
leave this for future exploration. At low redshift, LSST is not the most 
efficient SN search survey, and there will be an important role for 
$z<0.1$ surveys. Since SN at these redshifts are readily identifiable, this 
could be combined with the spectroscopic survey, in a next generation 
Nearby Supernova Factory. Thus we envision that a combination of 
the Nearby Supernova Factory$^+$, LSST, and possibly the Nancy 
Grace Roman Space Telescope or James Webb Space Telescope 
for NIR spectroscopy and possibly $z>1$ imaging could 
have important roles in $S^4$. 

An intriguing, but highly speculative, idea is whether sparseness in 
spectral wavelength could be added to sparseness in spectral phase 
and in redshift. Further study  on the spectral range 
necessary for effective Twins Embedding, or some additional spectral 
technique, would be welcome. 
Also, we do not currently understand why there is only 
0.02 mag internal dispersion in the 6600--7200 \AA\ restframe range, 
and where the additional external dispersion (e.g.\ up to 0.073 mag) 
comes from. If the external dispersion could be corrected, then 
each SN becomes even more powerful. 

$S^4$ demonstrates that a 
carefully crafted next generation supernova cosmology program 
could be highly illuminating for dark energy, while being 
observationally efficient.

\section{Conclusions} \label{sec:concl} 

Recent developments have opened the possibility that supernovae 
can be exceptionally well calibrated as distance indicators through 
measurement of their spectra at a single epoch near maximum light. 
This could lead to a highly time efficient survey with great dark energy 
sensitivity: one spectrum per SN, at a single epoch, with excellent 
statistical dispersion, and strong cosmological leverage from a 
small number of selected redshift slices. 

Planned wide field, deep, time domain surveys such as LSST will provide 
an extensive candidate list from imaging, then the best few thousand 
would be selected for spectrophotometric measurements following an 
optimized, sliced redshift distribution. Carefully crafted sparseness 
in redshift can be quite powerful and more efficient than a filled 
redshift range; we saw that it requires an increase of 25\% in the SN sample  
having the traditional even, rather than optimized, distribution to be essentially 
equivalent to the optimized result for the dark energy figure of merit. 

The dark energy figure of merit for this Sparse Spectroscopic 
Supernova Survey, $S^4$, approaches an impressive 300, 
assuming stringent systematics control is enabled by the 
recent advances in spectrophotometry, aided both by the 
demonstrated low 
dispersion and by the focused redshifts. 
We also assessed the impact 
of varying the low and high redshift ends, and the systematics 
level. A combination of LSST and Roman or JWST, and a low 
redshift survey such as a next generation Nearby Supernova 
Factory, would feed into and complement $S^4$. 
The outstanding sensitivity of supernovae for dark energy, and 
their systematics distinct from other probes, motivate a central 
role for them in next generation cosmology.

\acknowledgments 

I thank Greg Aldering, Alex Kim, and Saul Perlmutter for helpful 
discussions. 
This work is supported in part by the Energetic Cosmos Laboratory and by the 
U.S.\ Department of Energy, Office of Science, Office of High Energy 
Physics, under contract no.~DE-AC02-05CH11231.


\end{document}